\documentclass[12pt]{article}
\usepackage{amsmath}
\usepackage{amsfonts}   
\usepackage{amssymb}
\begin{document}
\date{}
\title{\textbf{Remarks on the Noncommutative Gravitational Quantum Well}}
\author{{Rabin Banerjee}\thanks{E-mail: rabin@bose.res.in}, \ {Binayak Dutta Roy}\thanks{E-mail: bnyk@bose.res.in} \ and {Saurav Samanta}\thanks{E-mail: saurav@bose.res.in}\\
\\\textit{S.~N.~Bose National Centre for Basic Sciences,}
\\\textit{JD Block, Sector III, Salt Lake, Kolkata-700098, India}}
\maketitle
                                                                                
\begin{quotation}
\noindent \normalsize 
A planar phase space having both position and momentum noncommutativity is defined in a more inclusive setting than that considered elsewhere. The dynamics of a particle in a gravitational quantum well in this space is studied. The use of the WKB approximation and the virial theorem enable analytic discussions on the effect of noncommutativity. Consistent results are obtained following either commutative space or noncommutative space descriptions. Comparison with recent experimental data with cold neutrons at Grenoble imposes an upper bound on the noncommutative parameter. Also, our results are compared with a recent numerical analysis of a similar problem. Finally, we provide a noncommutative version of the virial theorem for the case at hand.
\end{quotation}
\section{Introduction}
The subject of noncommutativity has a long history culminating in the paper by Snyder\cite{sny}. The hope was that noncommutative spaces would provide a natural background for a possible regularisation of quantum field theories\cite{sny,yang}. Since then different structures of noncommutative space have been considered\cite{ju, dou, majid}. Perhaps the most widely studied case is the canonical description; i. e. when the noncommutative parameter is a constant. Theories defined on such a noncommutative space are based on the Weyl-Wigner correspondence, in which all products are replaced by the star product.

A particularly interesting physical manifestation of canonical noncommutativity occurs in the context of the Landau problem-the planar motion of a charged particle subjected to a constant magnetic field\cite{nair,rab}. In refs. \cite{rab,sa} it was shown how noncommutativity could be shifted from the coordinates to the momenta and vice-versa. Also, the implications of noncommutativity in both phase space and configuration space variables were discussed.

In this paper we study the phenomenology of a quantum mechanical model with constant noncommutativity in both coordinates and momenta. This model is applied to the problem of a particle in the quantum well of the Earth's gravitational field to determine the effects of noncommutativity on the energy spectrum. Our analysis is totally analytical and compared with the numerical approach to a similar problem which has been done recently\cite{ba}. Finally we use the experimental results of \cite{nes,nes1} to put an upper bound on the noncommutative parameters. We also observe the asymmetric behaviour of the noncommutative parameters appearing among coordinates or momenta.

This paper is divided in five sections. In the next section we define the noncommutative space. A general phase space transformation is given that maps the noncommutative space variables with the commutative counterparts. Different parametrisations are discussed which reproduce the noncommutative algebra quoted in the literature\cite{rab,sa,ba}. We also show that, contrary to recent claims \cite{ba}, Planck's constant $\hbar$ need not be modified in order to simultaneously obtain noncommutativity in both coordinates and momenta. In section 3, the quantum gravitational well is introduced. After summarising the theoretical and experimental\cite{nes1} results of the energy spectrum in usual commutative space, we define the corresponding Hamiltonian in noncommutative space. The structure of this Hamiltonian is explicitly obtained in both noncommutative and commutative descriptions, leading to completely equivalent results. In section 4 the energy spectrum is computed by using the WKB approximation. Compatibility with the results obtained from the virial theorem is shown. An upper bound on the noncommutative parameter is derived by comparing with the recent experimental results given in \cite{nes1}. Concluding remarks and discussions are given in section 5. There are two appendices; in the first we discuss rotational symmetry on the noncommutative plane while in the second we derive a noncommutative version of the virial theorem for our problem.
\section{Noncommutative Phase Space}
In this section we describe the noncommutative phase space on which the model would be defined. Both configuration and momentum space noncommutativity are considered. It is generally believed\cite{ba} that this type of noncommutativity might lead to a redefinition of the Planck constant. However this is not mandatory, as our construction shows.

Consider a two dimensional space, where the position and momentum operators satisfy the standard Heisenberg algebra
\begin{eqnarray}
\begin{array}{rcl}
&&[x_i,x_j]=0\\
&&[p_i,p_j]=0\\
&&[x_i,p_j]=i\hbar\delta_{ij}.
\end{array}
\label{1}
\end{eqnarray}
Note that this algebra is invariant under the following symmetry transformation
\begin{eqnarray}
\begin{array}{rcl}
&&x_i\rightarrow p_i\\
&&p_i\rightarrow x_i\\
&&i\rightarrow -i.
\end{array}
\label{sym}
\end{eqnarray}
It is then possible to develop a coordinate space representation where the coordinates $x_i$ are diagonal and $p_i=-i\hbar\frac{\partial}{\partial x_i}$ or, alternatively, a momentum space description where the momenta $p_i$ are diagonal while $x_i=i\hbar\frac{\partial}{\partial p_i}$.

Now we take the following general phase space transformation
\begin{eqnarray}
&&y_i=x_i+\alpha_1\epsilon_{ij}p_j+\alpha_2\epsilon_{ij}x_j
\label{100}\\
&&q_i=p_i+\beta_1\epsilon_{ij}x_j+\beta_2\epsilon_{ij}p_j
\label{200}
\end{eqnarray}
where $\alpha, \ \beta$ are arbitrary constants. Here we enforce a symmetry leading to $y_i\rightarrow q_i$ and $q_i\rightarrow y_i$ under the transformation (\ref{sym}). Clearly this is possible if we introduce the following transformation
\begin{eqnarray}
&&\alpha_i\rightarrow \beta_i\\
&&\beta_i\rightarrow \alpha_i.
\end{eqnarray}
Thus a symmetry transformation, analogous to (\ref{sym}), in the modified $y-q$ plane is given by
\begin{eqnarray}
\begin{array}{rcl}
&&y_i\rightarrow q_i\\
&&q_i\rightarrow y_i\\ 
&&\alpha_i\rightarrow \beta_i\\
&&\beta_i\rightarrow \alpha_i\\
&&i\rightarrow -i.
\end{array}
\label{sym1}
\end{eqnarray}
Using (\ref{1}) we can show that the new coordinates $y$ and momenta $q$ satisfy the algebra
\begin{eqnarray}
&&[y_i,y_j]=-2i\hbar\alpha_1\epsilon_{ij}\\
&&[q_i,q_j]=2i\hbar\beta_1\epsilon_{ij}\\
&&[y_i,q_j]=i\hbar(1+\alpha_2\beta_2-\alpha_1\beta_1)\delta_{ij}+i\hbar(\alpha_2-\beta_2)\epsilon_{ij}.
\end{eqnarray}
Under the symmetry transformation (\ref{sym1}) the above algebra is invariant. 

So far we did not associate any specific values to the coefficients $\alpha$ and $\beta$'s. Now if we set
\begin{eqnarray*}
&&\alpha_1=-\frac{\theta}{2\hbar}\\
&&\beta_1=\frac{\eta}{2\hbar}\\
&&\alpha_2=\beta_2=0
\end{eqnarray*}
we obtain
\begin{eqnarray}
\begin{array}{rcl}
&&[y_i,y_j]=i\theta\epsilon_{ij}\\
&&[q_i,q_j]=i\eta\epsilon_{ij}\\
&&[y_i,q_j]=i(1+\frac{\theta\eta}{4\hbar^2})\hbar\delta_{ij}=i\hbar_{eff}\delta_{ij}
\end{array}
\label{3}
\end{eqnarray}
which reproduces the noncommutative structure given in \cite{ba}. The term $\frac{\theta\eta}{4\hbar^2}$ is interpreted\cite{ba} as the correction to the Planck constant. This correction is claimed as a consequence of noncommuting phase space algebra. As mentioned in \cite{zhang} it is possible to scale $\theta$ and $\eta$ in a proper way so that $\hbar_{eff}=\hbar$ but in that case the first two relations of eq. (\ref{3}) take rather complicated forms and the noncommutative parameters cannot be expressed simply as $\theta$ or $\eta$. However it has been shown by us \cite{rab,sa}, using a different approach that both $\theta$ and $\eta$ noncommutativity can be retained without altering the Planck constant. This is also achieved in the present context by taking the following values of $\alpha$ and $\beta$
\begin{eqnarray}
\begin{array}{rcl}
&&\alpha_1=-\frac{\theta}{2\hbar}\\
&&\beta_1=\frac{\eta}{2\hbar}\\
&&\alpha_2=\beta_2=\frac{1}{2\hbar}\sqrt{-\theta\eta}
\end{array}
\label{param}
\end{eqnarray}
which yields the noncommutative algebra
\begin{eqnarray}
\begin{array}{rcl}
&&[y_i,y_j]=i\theta\epsilon_{ij}\\
&&[q_i,q_j]=i\eta\epsilon_{ij}\\
&&[y_i,q_j]=i\hbar\delta_{ij}
\end{array}
\label{4}
\end{eqnarray}
so that the Planck constant is not modified. Physical applications of this type of noncommutative algebra may be found in \cite{rab,sa}.

It should however be mentioned that there exist studies \cite{bert} which show that the effect of a modified Planck constant was negligible with respect to the magnitude of the effect of interest. In this sense the modification (or otherwise) of the Planck constant is more of academic interest.

 The inverse phase space transformation is given by
\begin{eqnarray}
\begin{array}{rcl}
&&x_i=Ay_i+B\epsilon_{ij}y_j+Cq_i+D\epsilon_{ij}q_j\\
&&p_i=Ey_i+F\epsilon_{ij}y_j+Aq_i+B\epsilon_{ij}q_j
\end{array}
\label{5}
\end{eqnarray}
where
\begin{eqnarray}
&&A=\frac{2\hbar^2-\theta\eta}{2(\hbar^2-\theta\eta)}, \  \  \  \  \ B=-\frac{\hbar\sqrt{-\theta\eta}}{2(\hbar^2-\theta\eta)}\nonumber\\
&&C=\frac{\theta\sqrt{-\theta\eta}}{2(\hbar^2-\theta\eta)}, \  \  \  \  \ D=\frac{\theta\hbar}{2(\hbar^2-\theta\eta)}
\label{A}\\
&&E=-\frac{\eta\sqrt{-\theta\eta}}{2(\hbar^2-\theta\eta)}, \  \  \  \  \ F=-\frac{\hbar\eta}{2(\hbar^2-\theta\eta)}.\nonumber
\end{eqnarray}
Observe that $\theta$ and $\eta$ must have different signs so that the various coefficients are real and well defined which guarantees the hermeticity of physical operators $x, p$ and $y,q$.

\section{Gravitational Well}
The problem is first discussed in usual commutative space. We consider a two dimensional plane where a particle of mass $m$ is subjected to the Earth's gravitational field in one direction; the vertical taken to be described by the coordinate $x_1$. We assume that the gravitational acceleration $g$ is constant near the surface of the earth. The commutative Hamiltonian is given by
\begin{eqnarray}
H=\frac{1}{2m}(p_1^2+p_2^2)+mgx_1.
\label{ha}
\end{eqnarray}
Since the particle is free in the $x_2$ direction, its energy spectrum is continuous in that direction and the wave function can be written as
\begin{eqnarray}
\psi(x_2)=\int g(k)e^{ikx_2}dk.
\end{eqnarray}
In the other direction the wave function is the well known Airy function $\phi(\xi)$ with appropriate normalization\cite{lan},
\begin{eqnarray}
\psi_n(x_1)=A_n\phi(\xi)  \  ; \   \xi=\left(\frac{2m^2g}{\hbar^2}\right)^{\frac{1}{3}}(x_1-\frac{E_n}{mg}).
\label{psi1}
\end{eqnarray}
The zeroes of the Airy function, $\beta_n$ give the energy eigenvalues
\begin{eqnarray}
E_n=-\left(\frac{mg^2\hbar^2}{2}\right)^{\frac{1}{3}}\beta_n \ ; \ n=1, \ 2, \ 3...
\label{energy}
\end{eqnarray}
Below the classical turning point $x_n=\frac{E_n}{mg}$ the wave function oscillates and above $x_n$ it decays exponentially. This was observed experimentally by Nesvizhevsky {\it et al.}\cite{nes}. They form the ``well" by placing a horizontal reflecting mirror in the Earth's gravitational field. The neutron was used as the quantum particle since it is chargeless and has a longer life time ($\tau\simeq 885.7$s)\cite{par}. By placing an absorber above the mirror they allow a cold neutron beam to flow with a horizontal velocity $v_2=6.5{\textrm{ms}}^{-1}$ between the mirror and the absorber. Then they measure the number of transmitted neutrons as a function of absorber height: this was shown to be a step like function which manifests the quantum nature of the problem.

For theoretical computations, a more transparent and simple solution is obtained in the WKB approximation. The energy eigenvalues for the linear gravitational potential are given by,
\begin{eqnarray}
E_n&=&\left(\frac{9m}{8}[\pi\hbar g(n-\frac{1}{4})]^2\right)^{\frac{1}{3}}
\label{ener}\\
&=&\alpha_{n}g^{\frac{2}{3}} \ ; \ n=1, \ 2, \ 3...
\label{wkb}
\end{eqnarray}
where
\begin{eqnarray}
\alpha_n&=&\left(\frac{9m}{8}[\pi\hbar (n-\frac{1}{4})]^2\right)^{\frac{1}{3}}.
\end{eqnarray}
A summary of both theoretical and experimental results is given. Taking the values of constants as
\begin{eqnarray}
&&\hbar=\frac{1}{2\pi}({\textrm {Planck constant}})=10.59\times 10^{-35} \ {\textrm {Js}}\\
&&g={\textrm {gravitational acceleration}}=9.81 \ {\textrm {ms}}^{-2}\\
&&m={\textrm {mass of neutron}}=167.32\times 10^{-29} \ {\textrm {Kg}}
\end{eqnarray}
the first two energy levels found from (\ref{ener}) are\footnote{The error is $\sim \ 1\%$ compared to the results derived from (\ref{energy})},
\begin{eqnarray}
&&E_1=1.392 \ {\textrm {peV}}=2.23\times 10^{-31}{\textrm{J}}
\label{ener1}\\
&&E_2=2.447 \ {\textrm {peV}}=3.92\times 10^{-31}{\textrm{J}}.
\label{ener2}
\end{eqnarray}
From $E_1$ and $E_2$ the classical turning points are calculated to be
\begin{eqnarray}
&&x_1=\frac{E_1}{mg}=13.59 \mu{\textrm m}\\
&&x_1=\frac{E_1}{mg}=23.88 \mu{\textrm m}.
\end{eqnarray}
These are in reasonable agreement with the experimental results\cite{nes1}
\begin{eqnarray}
&&x_1^{\textrm {exp}}=12.2\pm1.8({\textrm {syst}})\pm0.7({\textrm {stat}}) \ (\mu{\textrm {m}})\\
&&x_2^{\textrm {exp}}=21.6\pm2.2({\textrm {syst}})\pm0.7({\textrm {stat}}) \ (\mu{\textrm {m}}).
\end{eqnarray}
Error bars for the above mentioned energy levels are
\begin{eqnarray}
&&\Delta E_1^{\textrm {exp}}=6.55\times 10^{-32} \ {\textrm{J}}=0.41 \ {\textrm {peV}},
\label{exp1}\\
&&\Delta E_2^{\textrm {exp}}=8.68\times 10^{-32} \ {\textrm{J}}=0.54 \ {\textrm {peV}}.
\label{exp2}
\end{eqnarray}
\subsection{Noncommutative Space Description}
The problem is next formulated on a noncommutative space. The analogue of the Hamiltonian (\ref{ha}) in noncommutative space is defined as
\begin{eqnarray}
H=\frac{1}{2m}(q_1^2+q_2^2)+mgy_1
\label{label}
\end{eqnarray}
where the variables satisfy the algebra (\ref{4}). To find the spectrum, two approaches are possible. One can directly work in the noncommutative space variables or use the phase space transformations to reduce the problem on the usual commutative space. We first discuss the second approach. Using the maps (\ref{100},\ref{200}) together with the parametrisation (\ref{param}), we find,
\begin{eqnarray}
H&=&\frac{1}{2m}(p_1^2+p_2^2)+mgx_1+\frac{\eta}{2m\hbar}\epsilon_{ij}p_ix_j+mg(-\frac{\theta}{2\hbar}p_2+\frac{\sqrt{-\theta\eta}}{2\hbar}x_2)\nonumber\\
&&+\frac{\eta^2}{8m\hbar^2}(x_1^2+x_2^2)+\frac{\eta\sqrt{-\theta\eta}}{8m\hbar^2}(x_ip_i+p_ix_i)-\frac{\theta\eta}{8m\hbar^2}(p_1^2+p_2^2).
\label{kunal}
\end{eqnarray}
Defining a new constant
\begin{eqnarray}
\gamma=\frac{2\hbar\theta}{4\hbar^2-\theta\eta}
\end{eqnarray}
and a new variable
\begin{eqnarray}
\bar{p_2}=p_2-m^2g\gamma
\end{eqnarray}
we can write the above Hamiltonian in the form
\begin{eqnarray}
H&=&\frac{1}{2m}(1-\frac{\theta\eta}{4\hbar^2})(p_1^2+\bar{p_2}^2)+\frac{\eta^2}{8m\hbar^2}(x_1^2+x_2^2)+\frac{\eta}{2m\hbar}(p_1x_2-\bar{p_2}x_1)\nonumber\\
&&+\frac{\eta\sqrt{-\theta\eta}}{8m\hbar^2}(x_1p_1+x_2\bar{p_2}+p_1x_1+\bar{p_2}x_2)\nonumber\\
&&+mg\{(1-\frac{\eta\gamma}{2\hbar})x_1+\frac{\sqrt{-\theta\eta}}{2\hbar}
(1+\frac{\gamma\eta}{2\hbar})x_2\}-\frac{m^3g^2\theta^2}{2(4\hbar^2-\theta\eta)}.
\label{newh}
\end{eqnarray}
Since the difference between $\bar{p_2}$ and $p_2$ is just a constant, they satisfy the same commutation relations. The eigenvalues of $\bar{p_2}$ are translated by an equal amount vis a vis those for $p_2$ and hence these are not distinguished. Also neglecting the additive constant in the Hamiltonian (\ref{newh}) we get
\begin{eqnarray}
H&=&\frac{1}{2m}(1-\frac{\theta\eta}{4\hbar^2})(p_1^2+p_2^2)+\frac{\eta\sqrt{-\theta\eta}}{8m\hbar^2}(x_ip_i+p_ix_i)+\frac{\eta}{2m\hbar}\epsilon_{ij}p_ix_j\nonumber\\
&&+\frac{\eta^2}{8m\hbar^2}(x_1^2+x_2^2)+mg\{(1-\frac{\eta\gamma}{2\hbar})x_1+\frac{\sqrt{-\theta\eta}}{2\hbar}
(1+\frac{\gamma\eta}{2\hbar})x_2\}.
\end{eqnarray}
This Hamiltonian is put in a more transparent form by first defining a modified gravitational acceleration $\tilde{g}$ in the following way
\begin{eqnarray}
&&m\tilde{g} \ {\textrm {cos}}\epsilon=mg(1-\frac{\eta\gamma}{2\hbar})\\
&&m\tilde{g} \ {\textrm {sin}}\epsilon=mg\frac{\sqrt{-\theta\eta}}{2\hbar}
(1+\frac{\gamma\eta}{2\hbar}).
\end{eqnarray}
The tilting angle with the $x_1$ axis is given by
\begin{eqnarray}
\epsilon={\textrm{tan}}^{-1}\frac{\sqrt{-\theta\eta}}{2\hbar}\left(\frac{2\hbar+\eta\gamma}{2\hbar-\eta\gamma}\right)
\label{ang}
\end{eqnarray}
while, 
\begin{eqnarray}
\tilde{g}=g\{(1-\frac{\eta\gamma}{2\hbar})^2-\frac{\theta\eta}{4\hbar^2}(1+\frac{\gamma\eta}{2\hbar})^2\}^{\frac{1}{2}}
\label{X}
\end{eqnarray}
Since the product $\theta\eta$ is negative, $\tilde{g}$ is always positive definite. Now we rotate in the $x_1-x_2$ plane by an angle $\epsilon$, so that the coordinate of a point in the rotated frame is given by
\begin{eqnarray}
\begin{array}{rcl}
&&x_{1}'={\textrm {cos}}\epsilon \  x_{1}+{\textrm {sin}}\epsilon  \ x_{2}\\
&&x_{2}'={\textrm {cos}}\epsilon \  x_{2}-{\textrm {sin}}\epsilon  \  x_{1}.
\end{array}
\label{i}
\end{eqnarray}
Correspondingly, the momenta are transformed :
\begin{eqnarray}
\begin{array}{rcl}
&&p_{1}'={\textrm {cos}}\epsilon \  p_{1}+{\textrm {sin}}\epsilon  \ p_{2}\\
&&p_{2}'={\textrm {cos}}\epsilon \  p_{2}-{\textrm {sin}}\epsilon \  p_{1}.
\end{array}
\label{ii}
\end{eqnarray}
Using (\ref{i}) and (\ref{ii}) it is easy to show that
\begin{eqnarray}
&&p_{1}'^{2}+p_{2}'^{2}=p_{1}^{2}+p_{2}^{2}
\label{iii}\\
&&x_{1}'p_{2}'-x_{2}'p_{1}'=x_{1}p_{2}-x_{2}p_{1}
\label{iv}\\
&&x_{1}'^{2}+x_{2}'^{2}=x_{1}^{2}+x_{2}^{2}
\label{v}\\
&&x'_ip'_i+p'_ix'_i=x_ip_i+p_ix_i.
\end{eqnarray}
Therefore in the rotated frame the noncommutative Hamiltonian is given by,
\begin{eqnarray}
H&=&\frac{1}{2m}(p_1'^2+p_2'^2)+m\tilde{g}x_1'+\frac{\eta}{2m\hbar}\epsilon_{ij}p_i'x_j'+\frac{\eta^2}{8m\hbar^2}(x_1'^2+x_2'^2)\nonumber\\
&&+\frac{\eta\sqrt{-\theta\eta}}{8m\hbar^2}(x_i'p_i'+p_i'x_i')
-\frac{\theta\eta}{8m\hbar^2}(p_1'^2+p_2'^2).
\label{223}
\end{eqnarray}
The primed and unprimed variables satisfy the same algebra; henceforth the primes are all dropped. Then we can identify the first three terms of the Hamiltonian (\ref{223}) exactly as the commutative Hamiltonian given in (\ref{ha}). This should be considered as the unperturbed Hamiltonian. The term $\frac{\eta}{2m\hbar}\epsilon_{ij}p_ix_j$ is effectively a Landau problem like term, where a magnetic field is present perpendicular to the $x_1-x_2$ plane. The term $\frac{\eta^2}{8m\hbar^2}(x_1^2+x_2^2)$ is practically an oscillating potential.

Since the noncommutative effects are rather small we first confine to the leading order approximation in $\theta$ and $\eta$. Moreover (\ref{X}) shows that in the leading order,
\begin{eqnarray}
\tilde{g}=g[1+O(\theta\eta)].
\end{eqnarray}
Hence the Hamiltonian (\ref{223}) in the first order approximation is given by, 
\begin{eqnarray}
H&=&\frac{1}{2m}(p_1^2+p_2^2)+mgx_1-\frac{\eta}{2m\hbar}(x_1p_2-x_2p_1)\\
&=&H_0-\frac{\eta}{2m\hbar}(x_1p_2-x_2p_1)
\label{hamil}
\end{eqnarray}
where $H_0$ is nothing but the commutative Hamiltonian already given in eq. (\ref{ha}). The energy spectrum pertaining to this Hamiltonian will be computed in section 4.
\subsection{Alternative Formulation}
Here we analyse the structure of the Hamiltonian directly in terms of noncommuting space variables. In the leading order approximation, (\ref{hamil}) shows that only the noncommutativity in the momentum variables is relevant. Thus to simplify matters, consider $\theta=0$ from the very beginning. It is then possible to provide a representation of the operators in the coordinate basis, such that the coordinates $y_i$ are diagonal while the momenta $q_i$ are given by the usual differential operator plus an extra piece,
\begin{eqnarray}
&&y_i\rightarrow y_i\\
&&q_i\rightarrow -i\hbar\frac{\partial}{\partial y_i}+\frac{\eta}{2\hbar}\epsilon_{ij}y_j.
\end{eqnarray}
It is simple to verify that this is a valid representation of the algebra (\ref{4}) with $\theta=0$. In this basis the Hamiltonian (\ref{label}) takes the differential form
\begin{eqnarray}
H=-\frac{\hbar^2}{2m}(\frac{\partial^2}{\partial y_1^2}+\frac{\partial^2}{\partial y_2^2})+\frac{\eta}{2m\hbar}\epsilon_{ij}y_j(-i\hbar\frac{\partial}{\partial y_i})+mgy_1+\frac{\eta^2}{8m\hbar^2}(y_1^2+y_2^2).
\label{Y1}
\end{eqnarray}
It is now possible to compare this with the expression given in (\ref{223}). Since (\ref{223}) involves commutative space variables, either the coordinate or momentum representation may be used. Specifically, in the standard coordinate representation one may replace $x_i'\rightarrow y_i$ and $p_i'\rightarrow -i\hbar\frac{\partial}{\partial y_i}$. Then for $\theta=0$ the result (\ref{Y1}) is reproduced. Obviously in the leading order approximation this leads to (\ref{hamil}).

It is straightforward to repeat the analysis for the case $\eta=0$. Expression (\ref{223}) reveals that there are no corrections and only the unperturbed Hamiltonian remains. Coming back to the noncommutative variable approach, the algebra (\ref{4}) implies that, for $\eta=0$, a momentum space representation can be constructed. This is given by,
\begin{eqnarray}
&&q_i\rightarrow q_i\\
&&y_i\rightarrow i\hbar\frac{\partial}{\partial q_i}-\frac{\theta}{2\hbar}\epsilon_{ij}q_j.
\end{eqnarray}
The Hamiltonian (\ref{label}) is then written as,
\begin{eqnarray}
H=\frac{1}{2m}(q_1^2+q_2^2)+mg(i\hbar\frac{\partial}{\partial q_1}-\frac{\theta}{2\hbar}q_2).
\label{kun}
\end{eqnarray}
This is next compared with (\ref{223}) for $\eta=0$. Apparently it seems that there is a mismatch. Using the standard momentum representation in (\ref{223}) so that $p_i'\rightarrow q_i$ and $x_1'\rightarrow i\hbar\frac{\partial}{\partial q_1}$ the results do not agree. However, as has been stressed even at the basic level by Dirac\cite{pdirac}, both coordinate and momentum representations admit a generalisation. This generalisation is, however, trivial since it can be absorbed in the phase of the wave function. In this problem we can set $p_i'\rightarrow q_i$ and $x_1'\rightarrow i\hbar\frac{\partial}{\partial q_1}-\frac{\theta}{2\hbar}q_2$ because the algebra remains unaffected. With this choice (\ref{223}) agrees with (\ref{kun}). Thus the commutative and noncommutative space descriptions are equivalent up to an irrelevant over all phase.

We conclude this section by observing that a direct comparison between commutative and noncommutative space descriptions is possible only if either $\theta=0$ or $\eta=0$. Then as can be seen from (\ref{4}), it is feasible to formulate the coordinate representation or the momentum representation, respectively. If both $\theta$ and $\eta$ are nonvanishing then it is somewhat problematic to construct an explicit representation directly in the noncommutative space. In that case the variable change approach leading to (\ref{kunal}) is necessary. 
\section{Bounds on Noncommutative Parameters}
Here the energy spectrum is computed and therefrom bounds on the noncommutative parameters are determined. Consider the Hamiltonian (\ref{hamil}) in the first order approximation. Now the term proportional to $\eta$ in the above Hamiltonian can be treated perturbatively. The unperturbed part $H_0$ is known to be exactly solvable in terms of Airy functions\cite{lan}. Furthermore, using the property that Airy function (or any bound state wavefunction vis a vis motion in the direction $x_1$) is real, it is easily seen that
\begin{eqnarray}
<p_1>_n=\int_0^{+\infty}dx_1\psi_n^*(-i\hbar\frac{\partial}{\partial x_1}\psi_n)=0.
\end{eqnarray}
This can also be understood physically from the fact that, for a bound state system, the average current flow in a particular direction is zero. So effectively the Hamiltonian turns out to be
\begin{eqnarray}
H=H_0-\frac{\eta}{2m\hbar}x_1p_2
\label{54}
\end{eqnarray}
In this way we see that, in the leading order, the noncommutative corrections are entirely encoded in the term
\begin{eqnarray}
H_I=-\frac{\eta}{2m\hbar}x_1p_2.
\end{eqnarray}
Such a correction term was also found in the approach of \cite{ba}. The energy corrections due to this term were calculated in \cite{ba} from the standard perturbation formula
\begin{eqnarray}
\Delta E_n=\int_0^{+\infty}dx_1\psi_n^*(x_1)H_I\psi_n(x_1).
\end{eqnarray}
where $\psi_n(x_1)$ is given in (\ref{psi1}). This integral was evaluated by using numerical methods to obtain the following corrections for the lowest two energy levels
\begin{eqnarray}
&&|\Delta E_1|=2.83\times 10^{29}\eta \  ({\textrm{J}})
\label{ar3}\\
&&|\Delta E_2|=4.94\times 10^{29}\eta \  ({\textrm{J}}).
\label{ar4}
\end{eqnarray}
Comparing the above result with (\ref{exp1}) and (\ref{exp2}) the upper bound obtained on the noncommutative parameter $\eta$ was found to be\cite{ba}, 
\begin{eqnarray}
&&|\eta|\lesssim 2.32\times 10^{-61} \ {\textrm {kg}}^2{\textrm {m}}^2{\textrm {s}}^{-2} \  \ (n=1)\\
&&|\eta|\lesssim 1.76\times 10^{-61} \ {\textrm {kg}}^2{\textrm {m}}^2{\textrm {s}}^{-2} \  \ (n=2).
\end{eqnarray}

In our approach we avoid the numerical analysis totally and follow the semi classical WKB method which works extremely well for a linear potential. We write the complete Hamiltonian (\ref{54}) in the form
\begin{eqnarray}
H&=&\frac{1}{2m}(p_1^2+p_2^2)+m(g-\frac{\eta}{2m^2\hbar}p_2)x_1
\label{Ha}\\
&=&\frac{1}{2m}(p_1^2+p_2^2)+mg'x_1
\end{eqnarray}
where $g'=g-\frac{\eta}{2m^2\hbar}p_2$. Since in the $x_2$ direction the particle is free, $p_2$ is a constant of motion. In the experiment painstakingly performed by Nesvizhevsky {\it et al.}\cite{nes} the expectation value of $p_2$ was
\begin{eqnarray}
<p_2>=10.91\times 10^{-27} \ {\textrm {Kg m s}}^{-1}.
\label{p2}
\end{eqnarray}
Now we can use (\ref{wkb}) to write the corrected energy values of the Hamiltonian (\ref{Ha}) as,
\begin{eqnarray}
E_n+\Delta E_n&=&\alpha_n(g')^{\frac{2}{3}}\nonumber\\
&=&\alpha_n(g-\frac{\eta}{2m^2\hbar}<p_2>)^{\frac{2}{3}}
\label{delt}
\end{eqnarray}
where $E_n$ corresponds to the unperturbed energy and $\Delta E_n$ is the correction. It is possible to find an analytic expression for $\Delta E_n$ from (\ref{delt}) by an expansion,
\begin{eqnarray}
E_n+\Delta E_n=\alpha_ng^{\frac{2}{3}}(1-\frac{\eta}{2gm^2\hbar}<p_2>)^{\frac{2}{3}}.
\end{eqnarray}
Retaining the leading $\eta$-order term we find,
\begin{eqnarray}
\Delta E_n=-\frac{\eta}{3gm^2\hbar}<p_2>E_n.
\label{bid}
\end{eqnarray}
As expected the same functional form of the result follows through the use of the virial theorem\cite{brau}. 

For a linear potential the virial theorem implies $<T>=\frac{1}{2}<V>$ where $T$ and $V$ are kinetic and potential energies, respectively. Then the total energy $E$ is given by $E=<T>+<V>=\frac{3}{2}<V>.$ In our problem $V=mgx_1$, so that $<x_1>=\frac{2E}{3mg}$. Now the perturbation term is $\Delta E=-\frac{\eta}{2m\hbar}<p_2><x_1>$. Hence we find
\begin{eqnarray}
\Delta E=-\frac{\eta E}{3m^2g\hbar}<p_2>
\end{eqnarray}
which reproduces the structure (\ref{bid}). Taking the values of $E_1$ and $E_2$ from (\ref{ener1}, \ref{ener2}) and $<p_2>$ from (\ref{p2}) we get on using (\ref{bid}),
\begin{eqnarray}
&&|\Delta E_1|=2.79\times 10^{29}\eta \ ({\textrm {J}})
\label{ar1}\\
&&|\Delta E_2|=4.90\times 10^{29}\eta \ ({\textrm {J}}).
\label{ar2}
\end{eqnarray}
Finally, using the experimental input from (\ref{exp1}, \ref{exp2}) leads to the following upper bounds on $\eta$;
\begin{eqnarray}
&&|\eta|\lesssim 2.35\times 10^{-61} \ {\textrm {kg}}^2{\textrm {m}}^2{\textrm {s}}^{-2} \  \ (n=1)\\
&&|\eta|\lesssim 1.77\times 10^{-61} \ {\textrm {kg}}^2{\textrm {m}}^2{\textrm {s}}^{-2} \  \ (n=2)
\end{eqnarray}
The energy corrections (\ref{ar1}, \ref{ar2}) are in excellent agreement with the numerical results (\ref{ar3}, \ref{ar4}) obtained by perturbing about the exact Airy function solutions. The same naturally holds true for the upper bound on $\eta$.
\section{Conclusions}
We have discussed a model of a particle in the quantum well of the Earth's gravitational field and a perfectly reflecting horizontal plane beneath, defined in a space with noncommuting coordinates and momenta. The energy spectrum in this model was computed analytically by exploiting the WKB approximation. Comparison with the experimental findings of \cite{nes,nes1} placed an upper bound on the $\eta$- noncommutativity parameter appearing in the algebra of momenta.

Our results were also in excellent agreement with the numerical calculations of \cite{ba} where a similar model was considered. However to put things in a proper perspective, we emphasise that there is a conceptual difference between our model (\ref{label}) and that of \cite{ba}. In our treatment the model is defined directly in the noncommutative space. It is then possible to analyse it either in terms of noncommuting variables (section 3.2) or, using the phase space transformations, in terms of commuting variables (section 3.1). The results agree in either formulation. In ref. \cite{ba}, on the contrary, the model was defined by taking the usual Hamiltonian (\ref{ha}) and exploiting the inverse phase space transformation (\ref{5}) to express it in terms of noncommuting variables. A perturbative expansion of the Hamiltonian in $\eta$ and $\theta$ is next carried out whereas we get a closed form expansion (\ref{223}) which truncates at the second order in the noncommutative parameters. For the leading order term, however, the two expressions match, upto a sign of $\eta$. This permitted us to make the necessary comparison.

An essential ingredient of this analysis was to exploit rotational symmetry to simplify the Hamiltonian. Normally this symmetry would be violated for constant noncommutativity. However, as shown in appendix 1 for the particular case of two dimensions, it holds.

Lastly, in appendix 2, we derive a noncommutative version of the virial theorem for the linear gravitational potential considered here. 
\section*{Appendix 1}
Here we discuss the validity of rotational symmetry in two dimensions. To see this take the following general structure for arbitrary n-dimensions,
\begin{eqnarray}
[y_i,y_j]=i\theta_{ij}, \  \ i,j=1,2,...n
\label{bha}
\end{eqnarray} 
Under rotations,
\begin{eqnarray}
\delta y_i=\omega_{ij}y_j \  \  \ ; \ \omega_{ij}=-\omega_{ji}.
\end{eqnarray}
Then taking the variation on both sides of (\ref{bha}),
\begin{eqnarray}
[\delta y_i,y_j]+[y_i,\delta y_j]=0
\end{eqnarray}
which implies,
\begin{eqnarray}
\omega_{ik}\theta_{kj}+\omega_{jk}\theta_{ik}=0.
\end{eqnarray}
This is not true in general. For $d=2$, we may write $\omega_{ij}=\omega\epsilon_{ij}, \ \theta_{ij}=\theta\epsilon_{ij}$ so that the above condition simplifies to,
\begin{eqnarray}
\theta\omega(\epsilon_{ik}\epsilon_{kj}+\epsilon_{jk}\epsilon_{ik})=0
\end{eqnarray}
which holds. This shows that rotational symmetry is valid in $d=2$ noncommutative space. For higher dimensions deformed transformations have to be appropriately defined which restore this symmetry\cite{wess,rabin,kk}.
\section*{Appendix 2}
A noncommutative version of the virial theorem in the present context is presented here. For a linear potential, the usual virial theorem states that the average kinetic energy of a particle is half of the average potential energy. In a noncommutative space this theorem gets modified. Using the Heisenberg equation of motion, for any arbitrary state we can write
\begin{eqnarray}
\frac{d}{dt}<y_iq_i>=\frac{1}{i\hbar}<[y_iq_i,H]>.
\label{anj}
\end{eqnarray}
where $y_i$, $q_i$ are the noncommutative space variables. For our case we take the Hamiltonian $H$ as given in (\ref{label}). For stationary states the left hand side of the above equation is zero. Using the explicit commutation relations given in (\ref{4}), the eq. (\ref{anj}) turns out to be
\begin{eqnarray}
2<\frac{q^2}{2m}>-<mgy_1>=-\frac{\eta}{m\hbar}<\epsilon_{ij}y_iq_j>+\frac{\theta}{\hbar}mg<q_2>.
\end{eqnarray}
This is the modified virial theorem where the corrections are found to be first order in $\theta$ and $\eta$. In the limit $\theta, \eta\rightarrow0$ the usual virial theorem is recovered.

\end{document}